\documentclass[10pt,letterpaper,aps,prl,reprint,superscriptaddress,showpacs,twocolumn,nofootinbib,floatfix,amsmath,amssymb]{revtex4-1}

\usepackage{graphicx}
\usepackage{amsmath,amssymb,mathrsfs}
\usepackage{bm}

\usepackage{enumitem}
\usepackage{color}
\usepackage{hyperref}
\usepackage{enumerate}
\usepackage{verbatim}

\usepackage{url}
\usepackage{breakurl}

\date{\today}

\def\be{\begin{equation}}
\def\ee{\end{equation}}

\DeclareMathOperator{\tr}{tr}

\begin{document}

\title{A Lower Bound on the Energy Density in Classical and Quantum Field Theories}
\author{Aron C. Wall}
\affiliation{
School of Natural Sciences,
Institute for Advanced Study,
Princeton, New Jersey 08540, USA}
\email{aroncwall@gmail.com}

\begin{abstract}
A novel method for deriving energy conditions in stable field theories is described.  In a local classical theory with one spatial dimension, a local energy condition always exists.  For a relativistic field theory, one obtains the dominant energy condition.  In a quantum field theory, there instead exists a quantum energy condition, i.e. a lower bound on the energy density that depends on information-theoretic quantities.  Some extensions to higher dimensions are briefly discussed.
\end{abstract}


\maketitle

\section{Introduction}

Perhaps the most important and powerful concept in physics is ``energy''.  In addition to being closely associated with time evolution and conservation laws, the energy must normally be bounded below, in order for a physical system to possess a stable ground state.

In a local field theory, the energy $H$ is the integral of a canonical energy density $h$, calculated using Noether's theorem.  In many classical field theories, $h$ is itself a positive quantity; for example in the Klein-Gordon theory, the Hamiltonian takes the form
$h = \tfrac{1}{2}(\partial_t \phi)^2 + \tfrac{1}{2}(\partial_i \phi)^2 \ge 0$.
However, in certain theories such as Maxwell, the canonical energy density is not gauge invariant or positive, and it is necessary to add an ``improvement term'' in order to obtain a suitable energy density \cite{belinfante1939spin,belinfante1940current,rosenfeld1979energy,bak1994energy}.  Since these improvement terms are total derivatives, they do not affect the integrated energy if the fields fall off fast enough at infinity.

In general relativity (GR), there is no covariant positive stress-tensor for the gravitational field itself, and the proof of the global positive energy theorem---which requires the matter fields to obey the dominant energy condition: $T_{ab}\,t^a u^b \ge 0$, for $t^a,\,u^a$ in the future lightcone---is more subtle \cite{schoen1979proof,schoen1981proof,witten1981new}.  However, the physical motivation for assuming that any particular energy condition holds generally is unclear \cite{Curiel:2014zba} (but cf. \cite{Parikh:2014mja})  Worse, quantum field theory (QFT) violates all positive energy conditions written in terms of local quantum fields \cite{epstein1965nonpositivity}.  Once these field theories are coupled to gravity, this raises grave questions \cite{Visser:1999de, Ford:2003qt} about whether global results such as singularity theorems \cite{penrose1965gravitational,hawking1965occurrence,Senovilla:2014gza} and theorems ruling out causality violations \cite{Tipler:1976bi,Hawking:1991nk,Morris:1988tu,Friedman:1993ty,Olum:1998mu,Visser:1998ua,Penrose:1993ud} still apply.  Yet some of these results can still be proven from plausible nonlocal inequalities, involving
integrals of the energy \cite{PhysRevD.17.2521,tipler_diff,borde87,Ford:1994bj,ForRom97,Ford:1997fa,ForRom99,Gao:2000ga,Graham:2007va} and/or entropic quantities associated with various regions of spacetime \cite{Cas08,BlaCas13,sing,BouFis15a}.

This letter outlines a general method for proving that stable local field theories will necessarily possess geometrically localized energy conditions.  That is, given the fact that an integrated energy density is positive, we will show that there always exists a lower bound on the energy density at any point.  The nature of this lower bound will depend on the dimension, and also whether the theory is classical or quantum.

For a classical theory with $d$ spatial dimensions, the lower bound on the energy density may depend on fields localized to a $d-1$ dimensional surface.  When $d = 1$, this proves the existence of a strictly local energy condition.  There is a potential loophole for gravitational theories, arising from the fact that the Hamiltonian energy density vanishes on-shell.  But the result still applies if one restricts attention to the matter sector, on any background spacetime admitting a positive global energy.

The quantum case is more subtle.  Here, the energy condition also depends on some purely information-theoretic quantities defined on \emph{one} side of the $d-1$ dimensional surface.  More precisely, the energy condition is ``semilocal'', meaning that it is invariant under all unitary transformations on one side (and does not depend at all on the other side).  A classic example of a semilocal quantity is the entanglement entropy of a region \cite{Sorkin83,BKLS86,Srednicki93,Calabrese:2004eu}.  As we shall see, the stability of a QFT is closely related to Strong Subadditivity \cite{lieb1973proof} of the entanglement entropy (which implies that the more strongly a quantum system $A$ is entangled with a system $B$, the less strongly it can be entangled with another system $C$).

In many (perhaps all) QFT's, this lower bound on the stress-energy tensor is given simply by a second derivative of the entanglement entropy. This suggests that a ``quantum dominant energy condition'' (QDEC) holds in every Lorentz-invariant QFT.  A special case---the quantum \emph{null} energy condition (QNEC)---has already been conjectured on the basis of quantum gravity arguments \cite{Wall:2011kb,BouFis15a}, and proven for conformal vacua \cite{Wall:2011kb}, free or superrenormalizable bosonic field theories \cite{BouFis15a,Bousso:2015wca} and in certain holographic contexts \cite{Koeller:2015qmn,Akers:2016ugt} (cf. \cite{Fu:2016avb})  However, this letter cannot rule out the possibility that in more complicated field theories, additional semilocal terms may need to be added to the QNEC to obtain the correct bound.

The core argument is simple yet novel.  Whenever you have a global energy condition, the knowledge of all the information in a given region places a lower bound on the energy in the complementary region.  Now when you learn \emph{more} about this quantity, its range of allowed values narrows and hence the lower bound is nondecreasing.  This basic logical truth turns out to imply the nontrivial field theory results stated above.

\section{Classical $d = 1$ Case}

Suppose we have a classical field theory with one spatial dimension, parameterized by some coordinate $x$, and a set of fields $\phi_1(x) \ldots \phi_n(x)$, optionally satisfying certain local constraints of the form
\be
C(\phi_i, \phi_i', \phi_i'' \ldots) = 0
\ee
involving some finite maximum number $D$ of $x$-derivatives.  The field data may be specified arbitrarily, as long as one satisfies these constraints, some choice of fall-off conditions at $x = \pm \infty$, and suitable differentiability conditions on the fields $\phi_i$.  It will be important that there are no \emph{nonlocal} constraints on the data.  We will not assume any kind of translation invariance, and everything that follows may be generalized to the case of a theory defined on a finite spatial interval in an obvious way.

In practice, such problems often arise when there is a 1+1 dimensional theory with a well-defined Cauchy problem, so that we may identify a 1 dimensional Cauchy slice $\Sigma$ (either spacelike or null) on which data may be specified, subject only to local constraints.  In this way the results that follow may be applied to field theories in 1+1 spacetime dimensions; however the time evolution of the initial data plays no role in what follows.

Let us assume that we can define an ``energy density'' $T(\phi_i, \phi_i', \phi_i'' \ldots)$, locally defined as a function of at most $K$ derivatives of $\phi_i$, which satisfies a \emph{global} or integrated energy condition:
\be\label{stable}
E = \int_{-\infty}^{+\infty} T\,dx \ge 0.
\ee
(This notation is schematic, covering many possible stability integrals such as
\be\label{sigma}
\int_\Sigma T_{ab}\, u^a d\Sigma^b \ge 0,
\ee
where $u^a$ is a vector in the future lightcone, $\Sigma$ is a Cauchy slice, and $d\Sigma^a$ is the natural integration measure for fluxes crossing $\Sigma$.)

Does any such theory \emph{also} obey a local energy condition?  The answer is yes.  However, the local energy condition may require the energy density to be improved by the addition of a total derivative term, i.e.~$T(x) + M'(x) \ge 0$, for some $M$.  This integrates to the same total energy $E$, so long as $M \to 0$ as $x \to \pm \infty$.

In order to derive the energy condition, we must identify the correct choice of $M$.  We illustrate our method with the following parable: suppose that an ant is marching along the $x$ axis from $x = -\infty$ towards $x = +\infty$, making note of all the field values she observes along the way.   Partway through the journey, having arrived at the point $x = x_0$, the ant asks herself, ``\emph{Given} everything I have observed so far (from $-\infty$ to $x_0$), what is the minimum possible energy I might encounter in the remaining part of my journey (from $x_0$ to $+\infty$)?''  Let us write this quantity as
\be
M(x_0) = \text{inf}\left( \int_{x_0}^{+\infty} T \, dx \,\middle|\, \phi_i(x < x_0) \right),
\ee
where the $\text{inf}(A|B)$ symbol means the lower bound on $A$, consistent with the knowledge in $B$.  The stability condition \eqref{stable} places a lower bound on $M$, ensuring it is well-defined (this is our sole use of global stability in this section).

Since the constraints and differentiability conditions are local, $M$ can depend only on the local field data at $x_0$.  The jump conditions constrain only $\text{max}(D-1, K-1)$ derivatives of $\phi_i$, so only that many derivatives of $\phi_i$ can appear in $M$.  Thus the ant does not actually require any long-term memory to calculate $M$; knowledge of a finite number of derivatives at the point $x_0$ suffices.

As the ant continues her journey to higher values of $x$, she learns more about the value of the fields, placing further constraints on the allowed states of the system.  Now the minimum value of any quantity cannot decrease upon learning more information.  Hence, for any point $x_1 = x_0 + \Delta x$ further along the journey ($\Delta x > 0$):
\begin{eqnarray}
M(x_0) \le
\text{inf}\left( \int_{x_0}^{+\infty} T \, dx \,\middle|\, \phi_i(x < x_1) \right) \\ =
\int_{x_0}^{x_1} T\,dx + M(x_1),
\end{eqnarray}
Thus, while $M$ can increase due to the ant learning more, it can only decrease when the ant actually passes some energy and leaves it behind.  By taking $\Delta x$ to be infinitesimal, one obtains the desired local inequality:
\be\label{mono}
T + M' \ge 0.
\ee
Thus, the theory obeys a local energy condition.

Now consider a relativistic field theory satisfying the stability condition \eqref{sigma}, with causal propagation of information, and no fluxes of 2-momentum through spatial infinity.  Let $\Sigma$ be a partial Cauchy slice extending from spacelike infinity to a point $x_0$.  Define $M_a(x_0)$ for null $a$ as the lower bound of $p_a$ in the complement to $\Sigma$, given the data on $\Sigma$.  Then $M_a(x_0)$ is well defined and satisfies a monotonicity property like \eqref{mono} if $x_0$ is shifted in a spacelike direction.  So we obtain a covariantly improved \emph{dominant} energy condition: $(T_{ab} + \epsilon^c_a \partial_c M_b)\,t^a u^b \ge 0$, where $t^a$ and $u^a$ point in the future lightcone.  (The improvement term is conserved on the first index, but is not manifestly symmetric.)

\section{Gravitational Theories}

The above result may be applied on any curved background spacetime, as long as it admits a positive global energy.

However, in diffeomorphism invariant theories such as GR (or dilaton gravity \cite{Grumiller:2002nm}, which is better behaved than GR when $d = 1$) the constraint equations imply that the Hamiltonian density vanishes on-shell, up to a total derivative term \cite{Arnowitt:1959ah}.  Thus, the argument of the previous section has a potential loophole: since $T$ is itself a total derivative, conceivably $T + M' = 0$, and $0 \ge 0$ is not an interesting energy condition!  But see \cite{Bartnik} for a construction that might work (outside event horizons).

\section{Quantum Field Theory}

Suppose we try to apply the same argument to QFT, where there are no local energy conditions \cite{epstein1965nonpositivity}.  Instead we obtain a \emph{semilocal} condition, which can depend on information-theoretic quantities on one side of the point.

Let us begin by considering a (possibly mixed) density matrix $\rho$ describing the state at a given moment of time in a 1+1 field theory.  This state may be restricted to any region $R$ to obtain $\rho_R$.  However, we cannot recover the full state just from knowing the states in $R$ and its complementary region $\overline{R}$, because we also have to know \emph{how} they are entangled.

We can still define the quantity $M$ as a bound on the expectation value of the energy on the right of a point $x_0$, given the density matrix $\rho_{x < x_0}$ in the causal wedge to the left of $x_0$:
\be
M(x_0) = \text{inf}\left( \int_{x_0}^{+\infty} \langle T \rangle \, dx \,\middle|\, \rho_{x < x_0} \right),
\ee
and it still follows that $\langle T \rangle + M' \ge 0$ at every point.

However, it is no longer the case that $M$ is a local function of the quantum fields in the vicinity of $x_0$.  In particular, there are various entropy inequalities; for example Strong Subadditivity \cite{lieb1973proof} states that for any 3 quantum subsystems $A,B,C$, the von Neumann entropy $S(R) = -\tr(\rho_R \ln \rho_R)$\footnote{For axiomatic characterizations of the von Neumann entropy, see \cite{AFN,ochs,wehrl,csiszar}.} satisfies
\begin{eqnarray}
S(AB) + S(BC) \ge S(ABC) + S(B), \label{SSA} \\
S(AB) + S(BC) \ge S(A) + S(C). \label{weak}
\end{eqnarray}
If we choose $B$ to be a neighborhood of the point $x_0$, and let $A$ and $C$ be the regions to the left and right of that neighborhood, then the second form of Strong Subadditivity \eqref{weak} implies that the \emph{more} strongly the local fields near $x_0$ are entangled with data to their left, the \emph{less} strongly they can be entangled with data to the right (and vice versa).

Thus, when stitching together the local density matrices into a consistent state of the whole system, there are nonlocal constraints.  Suppose we are handed some density matrices $\rho_A$, $\rho_B$, and $\rho_C$, and we ask whether we can find a consistent global state $\rho_{ABC}$ with finite energy that restricts to these states in the respective regions $A,B,C$.  Eq. \eqref{SSA} tells us that we might be able to find a consistent entangled state $\rho_{AB}$ and also a consistent state $\rho_{BC}$, yet be unable to combine them into a consistent state of the entire line $\rho_{ABC}$.  Because these constraints are nonlocal, $M$ can depend on information arbitrarily far to the left of the point $x_0$.  This explains how a QFT can obey global, but not local energy conditions.

This motivates us to identify
$M = -(\hbar/2\pi)S'(\rho_{x<x_0})$, so that
\be\label{QEC}
\langle T \rangle \ge \tfrac{\hbar}{2\pi}S'',
\ee
where $c = k_B = 1$.
Although $S$ is divergent, its derivatives are normally finite.\footnote{In nonsmooth conformal vacua, $T$ may diverge \cite{Davies:1977yv,ForRom99}, but $S''$ compensates \cite{Wall:2011kb}.  An additional $(S')^2$ term explains the limit on how far the negative and positive pulses may be separated.}
%
%
When evaluated along a null Cauchy slice, this inequality is called the QNEC, and was derived in special cases in \cite{Wall:2011kb,BouFis15a,Bousso:2015wca,Koeller:2015qmn,Akers:2016ugt}.  But we will argue that more generally,
\eqref{QEC}
probably holds more in all states of all Lorentz-invariant field theories, on all Cauchy slices.  In passing, we will rederive the fact that \eqref{QEC} is saturated for first order perturbations to the vacuum \cite{Wall:2009wm,Bianchi:2012br,Bousso:2015wca}.

We now justify our choice of $M$.
Although $M$ is nonlocal, it is still highly constrained.  For example, $M$ must be invariant under any unitary operator $U$ acting in a region strictly to the left of $x_0$:
\be
M(U \rho_{x < x_0} U^\dagger) = M(\rho_{x < x_0}),
\ee
because a unitary acting on the left does not change the set of states allowed on the right.  (Of course it is also unchanged by a unitary acting to the right of $x_0$, since by construction $M$ depends only on the left region $x < x_0$.)

This tells us that the dependence of $M$ on the physics to the left of $x_0$ must in a certain sense be purely information-theoretic; it is sensitive only to the entanglement of information, not to the details of the material in which the information is encoded.    However, $M$ may depend in a more detailed way on the physics right \emph{at} $x_0$, i.e. it need not be invariant under unitaries acting in an interval that includes $x_0$.  Let us refer to a functional of the density matrix satisfying these properties as \emph{(left) semilocal}.

More generally, we can imagine acting with a unitary that couples the region $x < x_0$ to some auxiliary system $Q$ whose initial state $\rho_{Q}$ is unentangled with the QFT state.  Such transformations are equivalent \cite{stinespring,choi} to trace-preserving, completely positive\footnote{A `completely positive' map is one which preserves positivity of the density matrix even if it is entangled with another system.} maps: $\rho \to F[\rho]$.  These transformations are usually not invertible and do not preserve the information to the left of $x_0$.  But by construction, this new state $F[\rho_{x < x_0}]$ must still be compatible with the state to the right of $x_0$.  Hence the lower bound on the energy to the right cannot increase under such a transformation (because we could imagine starting with a state arbitrarily near the lower bound, and then acting with $F$).
We conclude that $M$ is monotonically decreasing under such maps:
\be
M(F[\rho_{x < x_0}]) \le M(\rho_{x < x_0}).
\ee

We can also place a constraint on $M$ from dimensional analysis, assuming that $T$ (the energy density) is identified with the stress tensor (which is weight 2 in a 1+1 dimensional theory).  In a CFT, $M(x)$ must be weight 1 under scale-invariance.  Or, if we choose our Cauchy slice to be null, $M(v)$ must be weight 1 under a Lorentz boost $v \to av + b$.

Up to a multiplicative constant, the only entity I know of which satisfies all of these constraints is $-S'$.
This is indeed monotonically decreasing under all completely positive maps, as can be seen by replacing the derivative with a finite difference to get the conditional entropy on two subsystems $S(AB) - S(A)$.  By the first form of Strong Subadditivity \eqref{SSA} we now have
\begin{eqnarray}
S_{AB} - S_{A} \ge S_{ABQ} - S_{AQ}, \\
\mathrm{hence}\,\, S'(\rho_{x < x_0}) \ge S'(F[\rho_{x < x_0}])
\end{eqnarray}
Furthermore, in a relativistic theory, $S'$ places \emph{a} lower bound on the energy integral:
\be\label{GSL}
\frac{2\pi}{\hbar} \int_{x_0}^{+\infty} \langle T \rangle\,dx \ge -S'(\rho_{x > x_0}) \ge -S'(\rho_{x < x_0}),
\ee
where the first inequality follows \cite{Wall10,Wall11,Akers:2016ugt} from monotonicity of relative entropy \cite{araki1976relative,araki1977relative} together with the Unruh effect \cite{UW84,BW76},\footnote{Here $T$ must be the \emph{canonical} (Noetherian) energy density, with respect to which the vacuum state is thermal in $\int T (x - x_0)\,dx$ when restricted to the Rindler wedge $x > x_0$ \cite{Fursaev:1998hr}. Otherwise the addition of a local improvement term may also be necessary.} while
the second inequality follows from the second form of
Strong Subadditivity \eqref{weak}.

For a state that is a first order perturbation to the vacuum state ($\rho = \rho_0 + \delta\rho$), one can in fact prove that $M = -\hbar S'/2\pi$ (fixing the multiplicative constant): since both inequalities in \eqref{GSL} are saturated for $\rho_0$, and satisfied in all states, they must also be saturated at first order in $\delta \rho$ (cf. \cite{Wall:2009wm,Bianchi:2012br,Bousso:2015wca}).

It is hard to think of any other quantity which would satisfy the desired criteria.  For example we cannot substitute the derivative of the Renyi entropy $S_n \propto \ln \tr (\rho^{n})$, since this does not satisfy Strong Subadditivity.  More generally, let us suppose that the greatest lower bound is
\be
M = -\tfrac{\hbar}{2\pi} S' + G,
\ee
where $G$ is a positive (by \eqref{GSL}) semilocal quantity of weight 1, that vanishes for all first order peturbations to the vacuum. 
Then the semilocal quantum energy condition will take the form:
\be\label{QNEC}
\langle T \rangle \ge \tfrac{\hbar}{2\pi} S'' - G'.
\ee
Note that $G$ cannot be constructed out of any smooth functional of $S_n$ (including $S$) and their derivatives, because in order for the quantity to be weight 1, there would have to be a single derivative.  But in pure states,
$S_n(\rho_{x < x_0}) = S_n(\rho_{x > x_0})$, which implies that $S_n'$ is odd under spatial reflections, and hence cannot have a consistent sign.

Also, $G = 0$ whenever the QNEC has already been proved \cite{Wall:2011kb,BouFis15a,Bousso:2015wca,Koeller:2015qmn,Akers:2016ugt}, suggesting it vanishes generally.  That would imply that \eqref{GSL} is the strongest possible bound on the energy to the right of $x_0$.  If you start with an excited state, this tells you the maximum amount of energy extractable from that region (without using classical communication to teleport information \cite{Hotta:2010py,Hotta:2013rt,Verdon-Akzam:2015tma}).

Applying the same arguments covariantly using \eqref{sigma}, it is natural to conjecture a quantum dominant energy condition (QDEC):
\be\label{QDEC}
\langle T_{ab} \rangle \,t^a u^b \ge \tfrac{\hbar}{2\pi} \! \left(\epsilon^{c}_a \epsilon^{d}_b\,\partial_c \partial_d S \right) t^a u^b,
\ee
where $t^a$ and $u^a$ are restricted to the forwards lightcone.  (The QDEC implies the QNEC, by taking $t^a, u^a$ to be the same null vector.)  The improvement term is symmetric and conserved.

\section{Higher Dimensions}
We can also generalize the argument to the case of $d > 1$ spatial dimensions.  But now the analogue of $x_0$ will be a $d-1$ dimensional surface $\varsigma$, and it may be that $M$ depends in a nonlocal way on the fields on $\varsigma$.

We have a choice of which ``global'' energy condition to use.  One possibility is \eqref{sigma}, the positivity of the energy-momentum vector, found by integrating $T_{ab}$ along a Cauchy slice $\Sigma$.  A more interesting choice is the `average null energy condition' (ANEC), which states that for any null geodesic $\gamma$,
\be\label{ANEC}
\int_{-\infty}^\infty T_{vv}\, dv \ge 0,
\ee
where $v$ is an affine null parameter along $\gamma$.  (Recent arguments suggest that any reasonable QFT will satisfy the ANEC \cite{Hartman:2016lgu,Faulkner:2016mzt,Wall:2009wi,Hofman:2009ug}, cf. \cite{Kelly:2014mra,Verch:1999nt,Wald:1991xn,Fewster:2006uf,Kontou:2012ve,Kontou:2015yha}).

In either case, let us define $M(\varsigma)$ as the lower bound of the energy to the ``right'' side of $\varsigma$, given the state of all the fields on the ``left'' side.  We can now derive an energy condition at a point $p \in \varsigma$ in a similar manner as before.  We will do this by considering a variation $\delta \varsigma$ with support only in a small neighborhood of the point $p$.  By applying the same monotonicity argument as in the previous section, one can show that there exists a positive function of the form $T(p) + M'(\varsigma,\delta \varsigma) \ge 0$.  But now the improvement term $M$ might depend on fields anywhere along $\varsigma$!  So the improved energy condition might not be local, even classically.

In the case where we start with \eqref{ANEC} in a QFT, an argument similar to the previous section suggests that the QNEC \cite{BouFis15a} holds for $\varsigma$ lying on a stationary null surface:
\be
\langle T_{vv}(p) \rangle \ge \tfrac{\hbar}{2\pi {\cal A}} (\delta_v)^2 S(\varsigma)
\ee
where $v$ is an affine null coordinate and ${\cal A}$ is the transverse area along which the slice $\varsigma$ is translated by $\delta_v$.  But further work is needed to show that there is no additional correction term in general interacting field theories.

\begin{acknowledgements} \textit{Acknowledgements} I am grateful for conversations with Jason Koeller, Stefan Leichenauer, Zachary Fisher, Raphael Bousso, Maulik Parikh, William Donnelly, Xi Dong, Nima Arkani-Hamed, Netta Engelhardt, Don Marolf, Ted Jacobson, Juan Maldacena, Tom Hartman, Gary Horowitz, Tom Roman, and Erik Curiel, and for support from the Institute for Advanced Study, the Martin A. and Helen Chooljian Membership Fund, the Raymond and Beverly Sackler Foundation Fund, and NSF grant PHY-1314311. \end{acknowledgements}

\bibliographystyle{utphys}
\bibliography{all}

\end{document}